\documentclass[reprint,pre,amsmath,amssymb,aps]{revtex4-1}
\usepackage{amsmath,amssymb,amsfonts,graphicx,verbatim,hyperref,mathtools,bbold,float}
\usepackage{silence}
\usepackage{soul, color}
\usepackage{physics}
\usepackage[caption=false]{subfig}
\hypersetup{
	colorlinks=true,
	linkcolor=red,
	citecolor=blue,
}
\newcommand{\R}{\mathbb{R}}
\DeclareMathOperator\arctanh{arctanh}
\numberwithin{equation}{section}

\usepackage[dvipsnames,tables]{xcolor}

\newcommand{\m}{{\bf m}}
\newcommand{\h}{{\bf h}}
\newcommand{\sig}{{\boldsymbol\sigma}}
\newcommand{\Sig}{{\boldsymbol\Sigma}}
\newcommand{\thet}{{\boldsymbol\theta}}
\newcommand{\x}{\sig}
\newcommand{\C}{{\bf C}}
\newcommand{\J}{{\bf J}}
\newcommand{\Q}{{\bf Q}}
\newcommand{\X}{{\bf X}}
\DeclarePairedDelimiterX{\infdivx}[2]{(}{)}{%
  #1\;\delimsize\|\;#2%
}
\newcommand{\KLD}{D_{\mathrm{KL}}\infdivx}

\begin{document}

\title{Ensemble Inhibition and Excitation in the Human Cortex:\\
an Ising Model Analysis with Uncertainties}

\author{Cristian Zanoci}
\email{czanoci@mit.edu}
\affiliation{Department of Physics, Massachusetts Institute of Technology, 77 Massachusetts Avenue, Cambridge, MA 02139, USA}
\affiliation{Center for Brains, Minds and Machines, Massachusetts Institute of Technology, 77 Massachusetts Avenue, Cambridge, MA 02139, USA}
\author{Nima Dehghani}
\email{nima.dehghani@mit.edu}
\affiliation{Department of Physics, Massachusetts Institute of Technology, 77 Massachusetts Avenue, Cambridge, MA 02139, USA}
\affiliation{Center for Brains, Minds and Machines, Massachusetts Institute of Technology, 77 Massachusetts Avenue, Cambridge, MA 02139, USA}
\author{Max Tegmark}
\email{tegmark@mit.edu}
\affiliation{Department of Physics, Massachusetts Institute of Technology, 77 Massachusetts Avenue, Cambridge, MA 02139, USA}
\affiliation{Center for Brains, Minds and Machines, Massachusetts Institute of Technology, 77 Massachusetts Avenue, Cambridge, MA 02139, USA}

\begin{abstract}
The pairwise maximum entropy model, also known as the Ising model, has been widely used to analyze the collective activity of neurons. However, controversy persists in the literature about seemingly inconsistent findings, whose significance is unclear due to lack of reliable error estimates. We therefore develop a method for accurately estimating parameter uncertainty based on random walks in parameter space using adaptive Markov Chain Monte Carlo after the convergence of the main optimization algorithm. We apply our method to the spiking patterns of excitatory and inhibitory neurons recorded with multielectrode arrays in the human temporal cortex during the wake-sleep cycle. Our analysis shows that the Ising model captures neuronal collective behavior much better than the independent model during wakefulness, light sleep, and deep sleep when both excitatory (E) and inhibitory (I) neurons are modeled; ignoring the inhibitory effects of I-neurons dramatically overestimates synchrony among E-neurons. Furthermore, information-theoretic measures reveal that the Ising model explains about $80\%-95\%$ of the correlations, depending on sleep state and neuron type. Thermodynamic measures show signatures of criticality, although we take this with a grain of salt as it may be merely a reflection of long-range neural correlations.
\end{abstract}

\date{\today}


\maketitle
\section{Introduction}
\label{sec:intro}

One of the main questions in neuroscience is how to accurately model the dynamics of networks of firing neurons.
This question involves controversies not only about the correct dynamics, but also about the most relevant degrees of freedom in the network. One important example is the issue of temporal versus rate coding, i.e., whether the precise time of spiking or only overall spiking rates matter in the description of network dynamics. A number of theoretical \cite{Abeles1991corticonics,Diesmann1999Synchrony} and experimental \cite{Gray1989correlation,Abeles1993pattern,Riehle1997synch,shlens2006structure,Pillow2008spatiotemporal} studies provide evidence for the importance of the exact timing of spikes, but this view has been challenged by alternative perspectives advocating for less time-constrained and more probabilistic models \cite{Shadlen1998rate,Abbott1999variability,schneidman2006weak,Averbeck2006correlation,Josic2009stimulus}. Attempts at answering this central question have been hampered both by experimental difficulties in acquiring adequate data and 
computational challenges related to the exponential (in system size) growth of the number of dependencies that a network model has to capture. Recent advances in experimental techniques using multielectrode arrays \cite{Campbell_1991mea,Jones1992mea,Litke2004mea,Allen2018mea} enable us to simultaneously record the activity of large populations of neurons, further amplifying the need to formulate an effective theory describing the macroscopic characteristics of large neuronal networks given their numerous degrees of freedom. Statistical mechanics provides many examples of such theories that relate the macroscopic properties of matter to the interactions between its microscopic degrees of freedom.

One such family of models, known as maximum entropy models, consists of the least structured probability distributions that are consistent with a set of empirical statistics on finite data. In a landmark study, it was shown that pairwise maximum entropy models, also known as Ising models, based on average spiking probability and correlations between pairs of neurons, give a good description of the firing patterns in retinal ganglion cells \cite{schneidman2006weak}. Since then, these models have been widely used to describe the activity of ensembles of neurons in a variety of systems, both \textit{in vitro} and \textit{in vivo} \cite{shlens2006structure,shlens2009structure,yu2008small,tkacik2009spin,tkavcik2013simplest,tkavcik2014searching,ganmor2011sparse,ganmor2011architecture,tavoni2017functional}.

Despite Ising models' success in describing the statistics of spiking patterns, they also have certain limitations. First, it has been argued that higher-order neuron couplings could play an important role in population coding, so that pairwise couplings fail to capture the full dynamics \cite{Staude2010highorder,zylberberg2015input,montani2009impact,ohiorhenuan2010sparse, Macke2011input,montani2013statistical,koster2014modeling,shimazaki2015simultaneous,leen2015simple,zylberberg2015input}, especially if exact spike timing is important \cite{Martignon2000coding, Brown2004analysis}. Second, the model's reliability may be distance-dependent, leading to successful predictions for neurons separated by large distances, but poor fits to the activity of local clusters of neurons that might be dominated by high-order correlations due to distance dependent connectivity profiles \cite{ohiorhenuan2010sparse,Pernice2011structure,Yu2011high}. Finally, Ising models may not be scalable to the full size of real neuronal networks \cite{roudi2009pairwise,roudi2009statistical,roudi2009ising}.

Although pairwise maximum entropy models have known limitations and have been extensively studied before, they remain one of the few simple models that can explain the main characteristics of collective behavior. As new data sets for increasingly larger neuronal populations become available, it is imperative to rigorously test the applicability and predictive power of Ising models on these data sets. Unfortunately, without any approximations, the computational cost of making predictions using these models grows exponentially with the number of neurons, thus rendering them intractable \cite{ganmor2011architecture}. Moreover, in part due to these computational challenges, the existing literature on neural Ising models is largely devoid of any quantifications of uncertainties on their parameters, which makes it harder to resolve controversies about whether Ising models fit experimental data well \cite{schneidman2006weak,shlens2006structure} or not \cite{ohiorhenuan2010sparse,roudi2009pairwise}.
    
It is therefore timely to develop an improved method for neural Ising modeling that can be applied to modern experimental data sets and can quantify parameter uncertainties while remaining computationally tractable. We will introduce such a method in this paper, and then use it to study the collective behavior of cortical excitatory and inhibitory neurons during the wake-sleep cycle (wakefulness, light sleep, and deep sleep) at multiple timescales. We seek to identify differences between excitatory and inhibitory neurons, as well as their distinctive behavior during wakefulness and sleep. Finally, we will also study the thermodynamic properties of the learned models.

\section{Methods}
\label{sec:methods}

\subsection{Data}
\label{sec:data}
\begin{figure}
\centering
\includegraphics[width=\columnwidth]{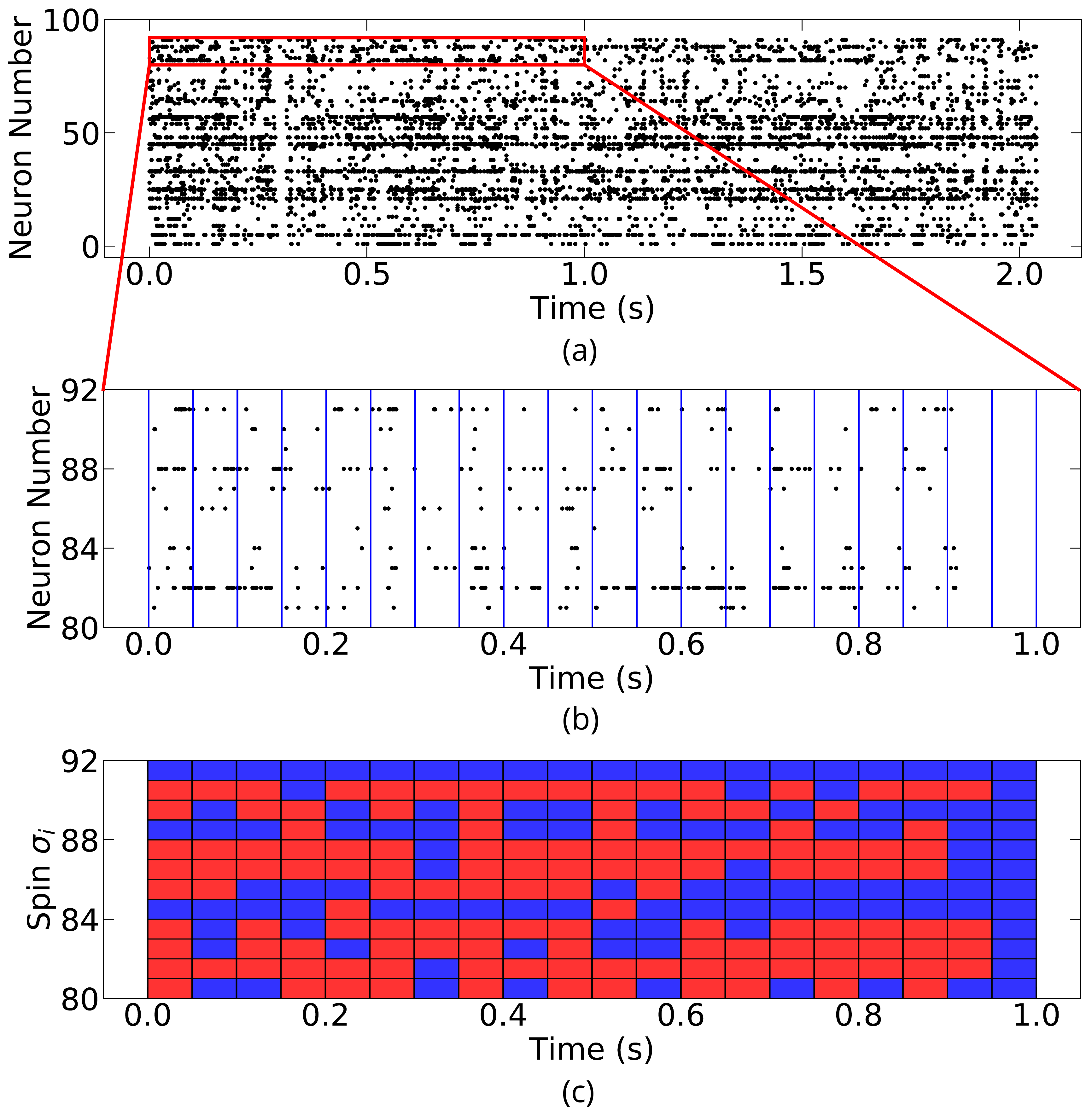}
\caption{\label{fig:raster} Schematic representation of experimental data. (a) The spike train response of a set of $92$ neurons in the human temporal cortex. (b) Discretization of the region delimited by the red rectangle into time bins of width $\Delta t = 50\text{ ms}$. (c) The raster corresponding to the discretization in panel (b), where red ($\sigma_i = +1$) denotes spiking of neuron $i$, and blue ($\sigma_i = -1$) represents silence.}
\end{figure}

We used data obtained from multielectrode recordings in layers II/III of the human temporal cortex. Data was initially sampled at $30$ kHz, then filtered and thresholded during the spike detection step. After spike sorting, a combination of morphological features of the spike waveforms along with the cross-correlogram of spike times were used to classify the cells as either excitatory (E) or inhibitory (I). This procedure produced a time-series of spike times for each of the $N$ neurons (Panel (a) of Figure~\ref{fig:raster}). The $12$ hour recordings spanning overnight sleep were staged, yielding multiple state labels: awake, light sleep (stages II-III), deep sleep (SWS; slow-wave sleep), and REM (rapid eye movement). In Sec.~\ref{sec:results}, we will use our method to analyze the awake, light sleep, and deep sleep states. Note that the data used in our analyses was devoid of any seizures. Additional details about the recordings and neuron classification procedure are presented in appendices \ref{sec:data-appendix} and \ref{sec:spike-appendix}. 

\subsection{Maximum-entropy models}
\label{sec:model}
The maximum-entropy concept originates from statistical mechanics, and the connection between maximum-entropy models and classical thermodynamic distributions has been long known \cite{jaynes1957information}. The main objective of maximum-entropy models is to construct a probability distribution that matches a set of empirical observations, but is otherwise as unconstrained and unstructured as possible. In this paper, we will focus on independent and pairwise models, which respectively match the first and second moments of the probability distribution of data.

Consider a network of $N$ neurons for which we  discretize each neuron's spike time-series into small windows of size $\Delta t$ (Panel (b) of Figure~\ref{fig:raster}). We performed this analysis for windows of size $20$, $50$, and $100\>$ms. For each neuron $i$, we assign $\sigma_i = +1$ if it generated an action potential (spiked) within this time window, or $\sigma_i = -1$ otherwise. Therefore, at each time step, our system can be described by a configuration vector $\x\in \{-1, 1\}^N$, visualized as a column of $N$ bits in Panel (c) of Figure~\ref{fig:raster}.

Suppose our data consists of a set of activity patterns $\X = \{\sig^{(1)}, \sig^{(2)}, \ldots, \sig^{(M)}\} \in \R^{N\cross M}$, where each $\sig^{(k)}$ is a configuration vector. Then we can compute the mean spiking probability of each neuron

\begin{equation}
m_i = \langle\sigma_i\rangle_{\X} = \dfrac{1}{M}\sum_{k=1}^M \sigma_i^{(k)},
\end{equation}
and the two-point function between pairs of neurons

\begin{equation}
Q_{ij} = \langle\sigma_i\sigma_j\rangle_{\X} = \dfrac{1}{M}\sum_{k=1}^M \sigma_i^{(k)}\sigma_j^{(k)}, 
\end{equation}
where $\langle \cdot \rangle_\X$ denotes the empirical average with respect to our data $\X$. We also define the covariance matrix 
\begin{equation}
C_{ij} = \langle\sigma_i\sigma_j\rangle_\X - \langle\sigma_i\rangle_\X\langle\sigma_j\rangle_\X= Q_{ij} - m_i m_j. 
\end{equation}

Our goal is to build a model for the observed probability distribution $P^{(N)}(\sig)$ which is consistent with the one- and two-point functions $\m$ and $\Q$ of the empirical data set $\X$.
More formally, we derive the probability distribution by maximizing its entropy, subject to the constraints that enforce agreement with $\m$ and $\Q$.
Using the method of Lagrange multipliers to impose each constraint, the solution to the optimization problem is given by the Boltzmann distribution \cite{jaynes1957information}:
\begin{equation}
\label{eq:boltzmann}
P^{(2)}(\sig,\thet) = \dfrac{e^{-\beta H(\sig,\thet)}}{Z(\thet)},
\end{equation}
where 
\begin{equation}
\label{eq:ham}
H(\sig,\thet) = -\sum_{i=1}^N h_i\sigma_i -\sum_{i, j = 1}^N J_{ij}\sigma_i\sigma_j = -\h^T\sig - \sig^T \J\sig,
\end{equation}
is the Hamiltonian of the system and 
$\thet = (\h,\J)$ is the parameter vector of our model. $P^{(2)}(\sig, \thet)$ is the probability that the network of neurons described by parameters $\thet$ is in a state $\sig$. The partition function $Z(\thet) = \sum_{\sig'}e^{-\beta H(\sig', \thet)}$ normalizes the probability distribution.
In physics applications, $\beta$ is interpreted as the inverse temperature, but in the context of neuroscience it is simply a parameter that scales $\h$ and $\J$, since the probability distribution depends on these parameters only through the combination $\beta \h$ and $\beta \J$. Therefore, without loss of generality, we can set $\beta = 1$ when fitting the model and absorb the scaling into our parameters $\h$ and $\J$. The dependence of the system on this scaling parameter $\beta$ will be explored in Section~\ref{sec:thermo}.

The Hamiltonian in Eq.~\eqref{eq:ham} represents an energy function that assigns a weight to each configuration of spikes and silences. This Hamiltonian is identical to that of an Ising model for a system of interacting spins \cite{kardar2007statistical}. Therefore, we will refer to the pairwise model as the {\it Ising model}. In contrast to the usual Ising models in physics, where couplings typically reflect translational symmetry in some number of dimensions, these neural Ising models allow arbitrary all-to-all couplings. The parameter vector $\h \in \R^N$ can be interpreted as the intrinsic tendency of each neuron to fire and $\J \in \R^{N\cross N}$ as the strength of pairwise interactions between the neurons. A positive $J_{ij}$ favors the neurons firing together, while a negative $J_{ij}$ does the opposite. We require that all the diagonal entries of $\J$ (i.e. self-interactions) are zeros ($J_{ii} = 0$), since $\sigma_i^2 = 1$ implies that $J_{ii}$ only contribute an irrelevant overall constant to our Hamiltonian. 
Without loss of generality, we take $\J$ to be symmetric ($J_{ij}=J_{ji})$, since $\sum_{ij} J_{ij}\sigma_i \sigma_j=\sum_{ij} J_{ij}\sigma_j \sigma_i=\sum_{ij} J_{ji}\sigma_i \sigma_j$.

Note that $\thet$ has $N(N+1)/2$ independent components, which for a system with $N = 92$ neurons yields a parameter space of size $4,278$. This is a reasonable number of parameters to describe our data set with $234,848$ entries (based on $50$ ms binning). However, if we include higher-order interactions, then the model would have at least $O(N^3)$ parameters, which will require significantly more data points to avoid overfitting.

A simplified version of the pairwise model is obtained by assuming that $\J = 0$, i.e.~ that each neuron spikes independently of all the others. This defines what we will refer to as the {\it independent model} $P^{(1)}(\sig, \h)$, which only constrains the mean spiking probability. Although the independent model is obviously not a realistic description of actual biological neural networks, it nonetheless provides a useful baseline comparison for the pairwise model. The main advantage of the independent model is that it is exactly solvable, since the partition function factorizes. One can easily show that its parameters must satisfy
\begin{equation}
\label{eq:h_in}
h_i = \arctanh(m_i).
\end{equation}
Unfortunately, such a closed-form correspondence between model parameters and data does not exist for the Ising model \cite{kardar2007statistical}. 

If we know the parameters $\thet$ of the Ising model, then we can compute any thermodynamic quantity, although it may take an exponential amount of time to evaluate the partition function $Z(\thet)$. In particular, we can compute the mean probability of spiking and the two-point correlation function:

\begin{equation}
\label{eq:m}
m_i(\thet) = \langle\sigma_i\rangle_\thet = \sum_{\sig} P^{(2)}(\sig, \thet) \sigma_i,
\end{equation}

\begin{equation}
\label{eq:C}
Q_{ij}(\thet) = \langle\sigma_i\sigma_j\rangle_\thet = \sum_{\sig} P^{(2)}(\sig, \thet) \sigma_i \sigma_j,
\end{equation}
where $\langle \cdot \rangle_\thet$ denotes the expectation value with respect to our model. This is known as the forward Ising problem. 

Our objective is to solve the inverse Ising problem - namely finding the best Ising model parameters $\thet$ that describe our data $\X$. Although this inference problem is complicated for large networks, efficient algorithms for solving it are an active area of research \cite{nguyen2017inverse}, and a variety of methods have been used over the years to learn maximum entropy models. These methods include histogram Monte Carlo \cite{broderick2007faster}, minimum probability flow \cite{sohl2011new}, adaptive cluster expansions \cite{cocco2009neuronal,cocco2011adaptive}, and pseudo-likelihood \cite{aurell2012inverse}. In this work, we use a combination of Markov Chain Monte Carlo (MCMC) \cite{metropolis1953equation,hastings1970monte} and gradient descent to iteratively estimate model averages of observables and update the parameters $\thet$. A detailed description of our algorithm is presented in Appendix~\ref{sec:learning}. 

Once we have learned the parameters of our model, a natural next step is to estimate the uncertainties associated with these parameters. As mentioned in Section~\ref{sec:intro}, this question has not been rigorously addressed in previous works on maximum entropy models. Our approach is to estimate the uncertainties using adaptive MCMC on the space of parameter vectors $\thet$. A complete description of our method is given in Appendix~\ref{sec:error}. One advantage of performing this random walk in parameter space is that we can fine-tune the solution previously obtained from the optimization algorithm.  

\subsection{Thermodynamic and \\information-theoretic quantities}
\label{sec:thermo}

The model parameters $\thet$ that we found can be interpreted as describing a system in thermal equilibrium at temperature $T = 1$, since as mentioned above, we set without loss of generality $\beta=1/k_B T=1/T=1$. By analogy with a statistical mechanics system described by a Boltzmann distribution, we can introduce the temperature $T$ as a scaling parameter of our Hamiltonian, which defines a one-parameter family of models whose thermodynamic properties can be explored. By varying $T$, we change the weights assigned to different spiking patterns. It is important to emphasize that $T$ is 
simply a model parameter, just like $\h$ and $\J$, and we have no actual neural network that corresponds to this model at $T\neq1$.
Further, we estimate the heat capacity $C(T)$ and entropy $S(T)$ of our network. The heat capacity of a network of neurons can be interpreted as the variance of the surprise, where the surprise $-\log P(\sig)$ determines how unexpected a particular spiking pattern $\sig$ is for the network \cite{tkavcik2014searching,tkavcik2015thermodynamics,mora2015dynamical}. A small heat capacity indicates that all spiking patterns appear with roughly the same probability, whereas a large heat capacity suggests that there is a balance between a few frequent patterns and multiple rare patterns \cite{mora2011biological,mora2015dynamical}. Moreover, divergences in the heat capacity can be used to determine the presence of a critical point. 

The entropy $S$, on the other hand, can be used to compute the effective number $2^S$ of spiking patterns of our system, which is an indicator of the size of the neural vocabulary. It also provides a bound on the network's capacity to encode and transmit information \cite{tkavcik2014searching}. Furthermore, we can use the information-theoretic interpretation of entropy to asses the accuracy of our model. For this, we define a hierarchy of models, consisting of the independent $P^{(1)}(\sig, \h)$, pairwise $P^{(2)}(\sig, \thet)$, and observed $P^{(N)}(\sig)$ models, in this order. Each subsequent model captures more correlations among the data, with the empirical model capturing all the correlations. Given how the models are sorted from least to most structured, their respective entropies should satisfy $S_1\geq S_2 \geq S_N$. The amount of correlation in the network is quantified by the ``multi-information''

\begin{equation}
\label{MultiInfoEq}
I_N\equiv S_1 - S_N,
\end{equation} 
i.e., the decrease in entropy relative to the independent model \cite{cover2012elements,schneidman2003network}. Similarly, $I_2 \equiv S_1 - S_2$ measures the decrease in entropy that is solely due to pairwise correlations. Therefore, the multi-information ratio $I_2 / I_N$ can be used to quantify the fraction of correlations captured by the pairwise model \cite{schneidman2006weak,mora2011biological}.

To compute entropies and heat capacities, we take the following steps. For the empirical entropy $S_N$ we use a low-bias estimator, specifically the Bayesian estimator with a centered Dirichlet mixture as its prior \cite{archer2013bayesian}. The entropy $S_1$ of the independent model can be computed analytically \cite{kardar2007statistical}. As for the entropy of the pairwise model, it is not feasible to compute the probability associated with every spiking pattern, and even Monte Carlo sampling would lead to a poor approximation of the probability distribution due to the exponential growth of the phase space. A standard technique for approximating the entropy relies on integrating the heat capacity \cite{tkavcik2015thermodynamics, tkavcik2014searching}. However, this method requires generating Monte Carlo samples at many intermediate temperatures in order to get an accurate estimate of the integral, which can get computationally expensive. For our purposes, we therefore choose to use the Wang-Landau algorithm \cite{wang2001efficient}, which is better suited for this task and yields an estimate for both the entropy and the heat capacity. A summary of the algorithm is given in Appendix~\ref{sec:wang-landau}.
\section{Results}
\label{sec:results}

In this section, we apply our methods to the above-mentioned data from \textit{in vivo} multielectrode array recordings of neurons in the human temporal cortex. We construct maximum-entropy models of both inhibitory and excitatory neurons across multiple sleep stages. The figures presented in this section are based on the temporal bin size $\Delta t=50$ ms, thus striking a good balance between capturing correlations among neurons and providing enough data for analysis. This value of $\Delta t$ is slightly larger than the conventional $20$ ms window used for retinal neurons \cite{schneidman2006weak,ganmor2011architecture,tkavcik2014searching,ioffe2017structured}, reflecting the sparsity of activity patterns in the temporal cortex \cite{nghiem2018maximum}. We repeated our analysis for both smaller ($\Delta t = 20 \text{ ms}$) and larger ($\Delta t = 100 \text{ ms}$) time bins, and confirmed that our conclusions also hold on these time scales.

\begin{figure*}
\centering
\includegraphics[width=\textwidth]{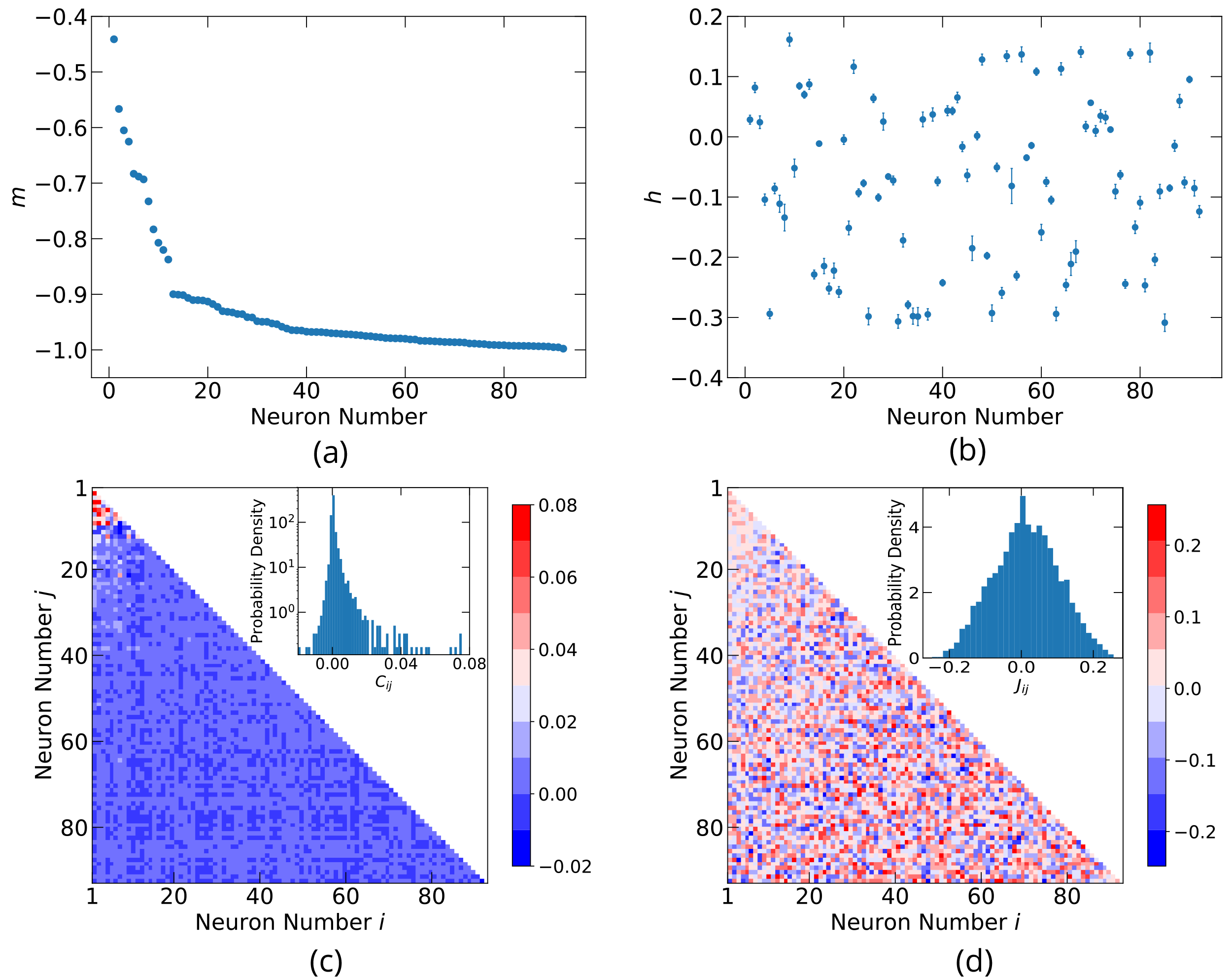}
\caption{\label{fig:obs} Maximum entropy model for the population of $92$ neurons in the awake state. (a) The mean probability of spiking $m_i = \langle \sigma_i \rangle$. (b) The bias terms $h_i$ of the inferred model, with a mean error of $0.01$. (c) The correlation coefficients between pairs of neurons $C_{ij} = \langle \sigma_i\sigma_j \rangle - \langle \sigma_i \rangle\langle \sigma_j \rangle$. The inset shows the population distribution of the correlation coefficients. (d) The pairwise coupling terms $J_{ij}$ of the inferred model, with a mean error of $0.004$. The inset shows the population distribution of the pairwise couplings. Neurons are ordered by decreasing spiking frequency.}
\end{figure*}

Figure~\ref{fig:obs} shows a sample fit of the pairwise model to the data recorded from $92$ neurons in the human temporal cortex during wakefulness, with the neurons sorted in order of decreasing activity $m_i$. Panels (a) and (c) display the measured statistical properties $\m$ and $\C$ of the network, while panels (b) and (d) show the inferred model parameters $\h$ and $\J$. This analysis shows that the majority of neurons have a very low probability of spiking, with the last $65$ neurons firing in less than $3\%$ of all time bins. The values of the covariance $C_{ij}$ for these less active neurons are close to zero, reflecting the fact that pairs of neurons are likely to be simultaneously silent. The majority of bias terms $h_i$ take on negative values, thus showing the neurons' intrinsic tendency to remain silent. 

The model uncertainties on $\h$ are slightly larger than those on $\J$, but still below $5\%$. The couplings $J_{ij}$ between neurons are widespread and can have either sign. The distribution of $J_{ij}$ is seen to be roughly symmetric and centered around zero. This behavior is reminiscent of spin glasses \cite{edwards1975theory}, where competing interactions lead to frustration in the system. Therefore, we would expect the high-dimensional energy landscape to become increasingly uneven and develop many local minima \cite{edwards1975theory,tkavcik2014searching}. This is consistent with the fact that multiple microscopic realizations of a system can lead to very similar macroscopic behaviors. We find that qualitatively similar observations and conclusions apply to both light and deep sleep states.

\subsection{Reliable neurons}
\label{sec:reliable}

\begin{figure}
\centering
\includegraphics[width=\columnwidth]{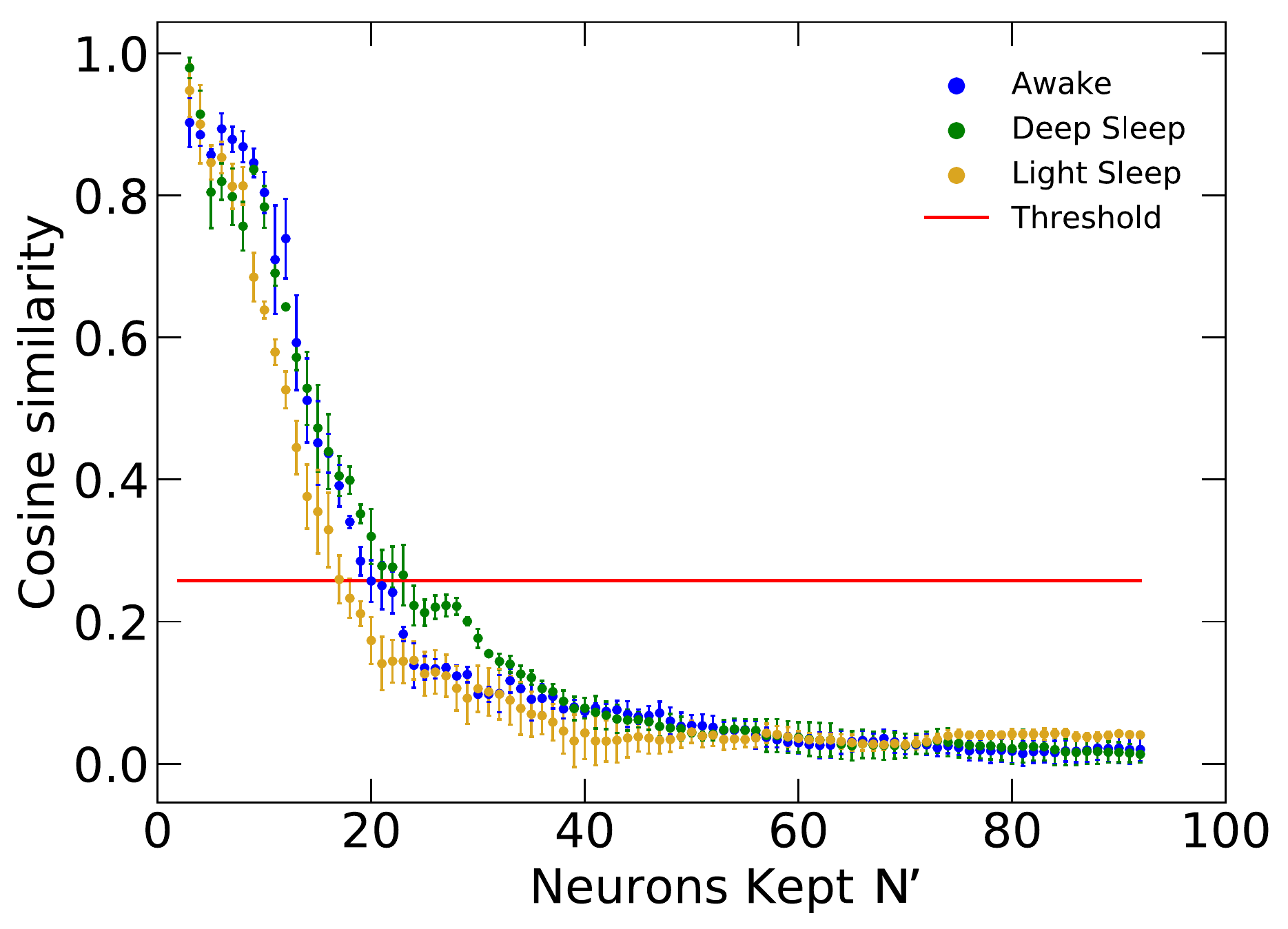}
\caption{\label{fig:cosine} Cosine similarity between all pairs of $10$ parameter vectors $\thet$ estimated from $10$ independent runs of our algorithm, when keeping only the $N'$ most active neurons.
Higher values of the cosine similarity indicate a smaller variance in parameters from run to run, so we see that the parameters of more active neurons can be more reliably measured. Our threshold for reliable neurons (horizontal line) is set at $0.25$.}
\end{figure}

We confirm numerically that the optimization problem does indeed have multiple local minima. By running the algorithm described in Appendix~\ref{sec:learning} with different initializations, we observe that the optimization subroutine converges to a slightly different set of parameters each time. Since we want to be able to meaningfully interpret the model parameters $\thet$, we will now investigate this issue.

To do so, we introduce cosine similarity as a measure of how reliably the parameters $\thet$ are replicated from run to run. The cosine similarity between two vectors $\thet^{(i)}$ and $\thet^{(j)}$ is simply the cosine of the angle between them, computed via their dot product
\begin{equation}
\label{eq:cosine}
\cos(\thet^{(i)}, \thet^{(j)}) \equiv \dfrac{\thet^{(i)} \cdot \thet^{(j)}}{|\thet^{(i)}| |\thet^{(j)}|},
\end{equation}
where $|\thet^{(i)}|\equiv(\thet^{(i)}\cdot\thet^{(i)})^{1/2}$ denotes vector length. If two parameter vectors differ only slightly, then they are almost aligned in the parameter space and their cosine similarity is close to $1$. If instead the difference between parameters is large, the two vectors will be roughly perpendicular in their high-dimensional parameter space and their cosine similarity will be near zero.

There are two candidate explanations for why our algorithm might produce different parameters $\thet$ for different initializations: 
\begin{enumerate}
    \item The hypothetical true values of $\m$ and $\C$ that we would measure if we had access to infinite data unfortunately lead to multiple local optima when fitting for $\thet$. 
    \item These true $\m$ and $\C$ would give a unique local and global optimum $\thet$, but the empirical $\m$ and $\C$ that we estimate from our finite data $\X$ are sufficiently far from the true values, so that local optima arise.
\end{enumerate}
The more two neurons $i$ and $j$ spike, the more accurately and reliably we can estimate $m_i$ and $J_{ij}$, whose uncertainties scale roughly as the square root of the number of spikes.
To distinguish between explanations $1$ and $2$, we therefore perform $10$ independent estimations of $\thet$ after discarding all but the $N'$ most active neurons, and plot the average cosine similarity between pairs of parameter vectors as a function of $N'$ in Figure~\ref{fig:cosine}. The results support the second hypothesis: parameters inferred from only the most active neurons are quite reliably recovered multiple times with different initializations, whereas those involving less active neurons are not. 

\begin{figure}
\centering
\includegraphics[width=\columnwidth]{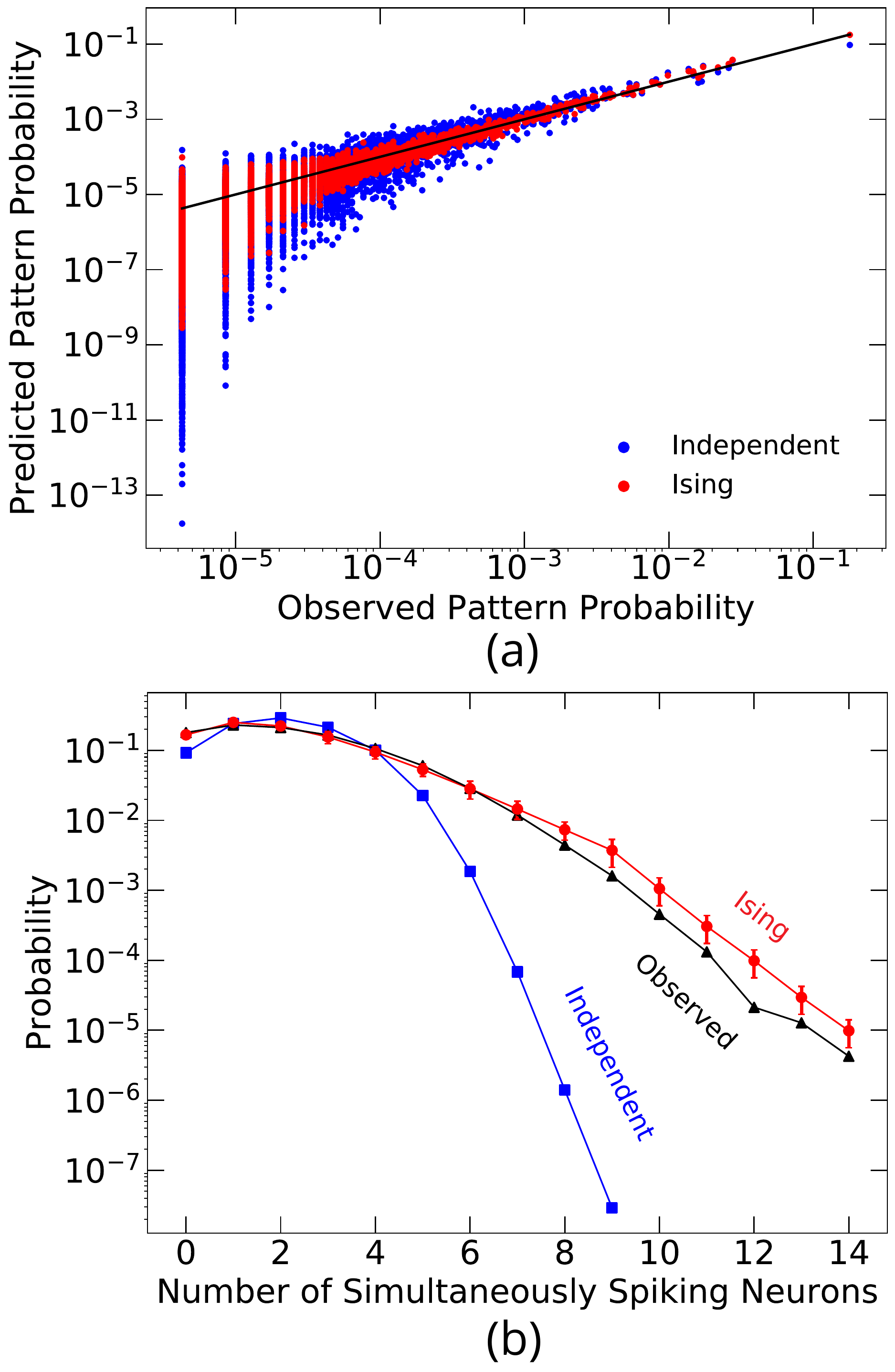}
\caption{\label{fig:good_model} 
Model predictions for the network of $19$ reliable neurons during the awake state. (a) The probability of occurrence of each firing pattern predicted by the maximum entropy model ($P^{(1)}(\sig, \h)$ or $P^{(2)}(\sig, \thet)$) is plotted against the observed pattern frequency from measured data ($P^{(N)}(\sig)$). 
The black line corresponds to prediction matching observation. (b) The predicted and observed distribution of the number of simultaneously spiking neurons in each time bin. Error bars are asymmetrical because of the logarithmic scale. Similar results hold across all states and time binnings.}
\end{figure}

\begin{table*}
\begin{center}
\begin{tabular}{l|l|r|c|c|c|c|c|c} 
 \hline
 \hline
 Neuron& State & $N$&\multicolumn{2}{c|}{KL divergence per neuron}& $S_1/N$ & $S_2/N$ & $S_N/N$ & $I_2/I_N$ \\ 
type & & & Independent & Ising & & & &\\ 
 \hline
 I & Awake & 23 & $0.0146(1)$ & $0.0044(1)$& $0.3326(1)$ & $0.3227(2)$&	$0.3204(2)$&	$0.81(2)$\\ 
 I & Light Sleep & 23 & $0.0325(1)$&	$0.0102(1)$&	$0.3578(1)$&	$0.3337(2)$&	$0.3282(2)$&	$0.81(1)$ \\ 
 I & Deep Sleep & 23 & $0.0376(1)$ &	$0.0129(1)$&	$0.3611(3)$ &	$0.3347(3)$ &	$0.3292(3)$ &	$0.83(2)$ \\
 E & Awake & 6 & $0.0008(2)$&	$0.00010(3)$&	$0.3975(3)$ &	$0.3893(3)$&	$0.3890(3)$&	$0.96(7)$ \\ 
 E & Light Sleep & 6 & $0.0010(2)$&	$0.00030(3)$&	$0.3205(3)$&	$0.3102(2)$&	$0.3092(3)$&	$0.91(5)$ \\ 
 E & Deep Sleep & 14 & $0.0054(1)$&	$0.0016(1)$&	$0.2885(1)$&	$0.2858(2)$&	$0.2851(2)$&	$0.80(8)$ \\
 I and E & Awake & 19 & $0.0185(1)$&	$0.0056(1)$&	$0.4826(2)$ &	$0.4698(2)$&	$0.4673(2)$&	$0.84(2)$ \\ 
 I and E & Light Sleep & 16 & $0.0323(1)$&	$0.0116(1)$&	$0.3976(1)$&	$0.3811(2)$&	$0.3792(1)$&	$0.90(1)$ \\ 
 I and E & Deep Sleep & 23 & $0.0444(1)$&	$0.0189(1)$&	$0.4374(2)$&	$0.4055(3)$&	$0.4033(3)$&	$0.94(1)$ \\
 \hline
 \hline
\end{tabular}
\end{center}
\caption{
Information-theoretic quantities for populations of reliable inhibitory and excitatory neurons across different sleep states. The digits in parenthesis represent the uncertainty in the last digit. 
}
\label{table:info}
\end{table*}

\begin{figure*}
\centering
\includegraphics[width=\textwidth]{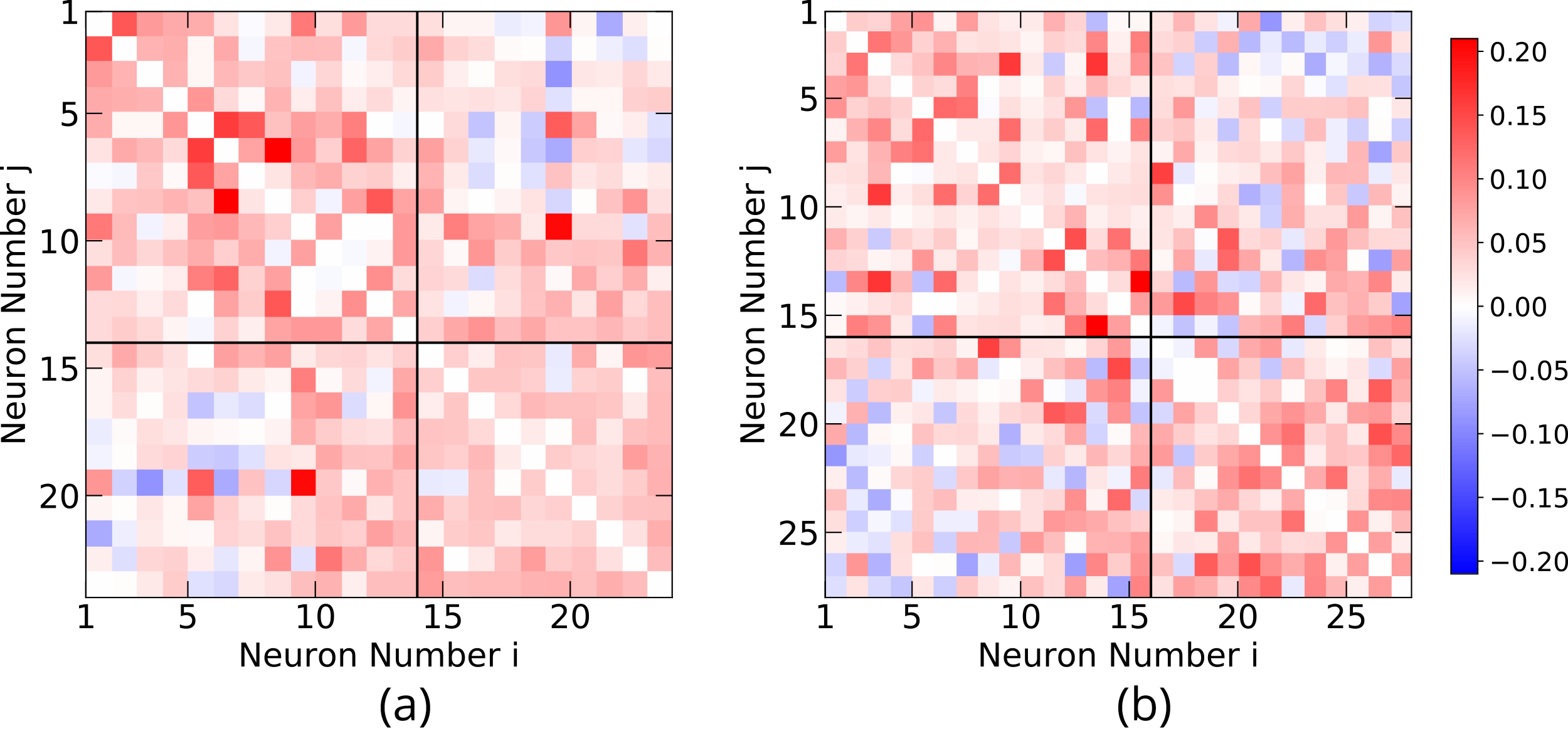}
\caption{\label{fig:EI} The pairwise coupling matrix $J$ inferred from the population of reliable neurons in the deep sleep state. Neurons are sorted by type, with (a) inhibitory neurons at positions $1-13$ and excitatory neurons at positions $14-23$ for patient $1$, (b) inhibitory neurons at positions $1-15$ and excitatory neurons at positions $16-27$ for patient $2$. The couplings $J_{ij}$ among inhibitory neurons, as well as the couplings among excitatory neurons (diagonal blocks) are almost entirely positive. The couplings $J_{ij}$ between inhibitory and excitatory neurons (off-diagonal blocks) display a mix of both positive and negative values.}
\end{figure*}

We therefore define \textit{reliable neurons} to be those for which the cosine similarity is above a given threshold, set here to $0.25$. Intuitively, reliable neurons are those for which we have enough data to confidently infer their model parameters. For our data, this criterion corresponds to neurons firing in at least $5\%$ of the time windows, and selects approximately $20$ neurons as reliable for each sleep state, roughly equally split between inhibitory and excitatory neurons. In the remainder of this section, we will only consider reliable neurons. 

In a related approach based on reliable interactions \cite{ganmor2011sparse}, during the model fitting, only the frequent activity patterns of the network were taken into account and all the configurations whose occurrence rate was below a certain threshold were discarded. Our approach is similar, except that instead of discarding time segments, we discard the least active neurons. 

\begin{figure*}
\centering
\includegraphics[width=\textwidth]{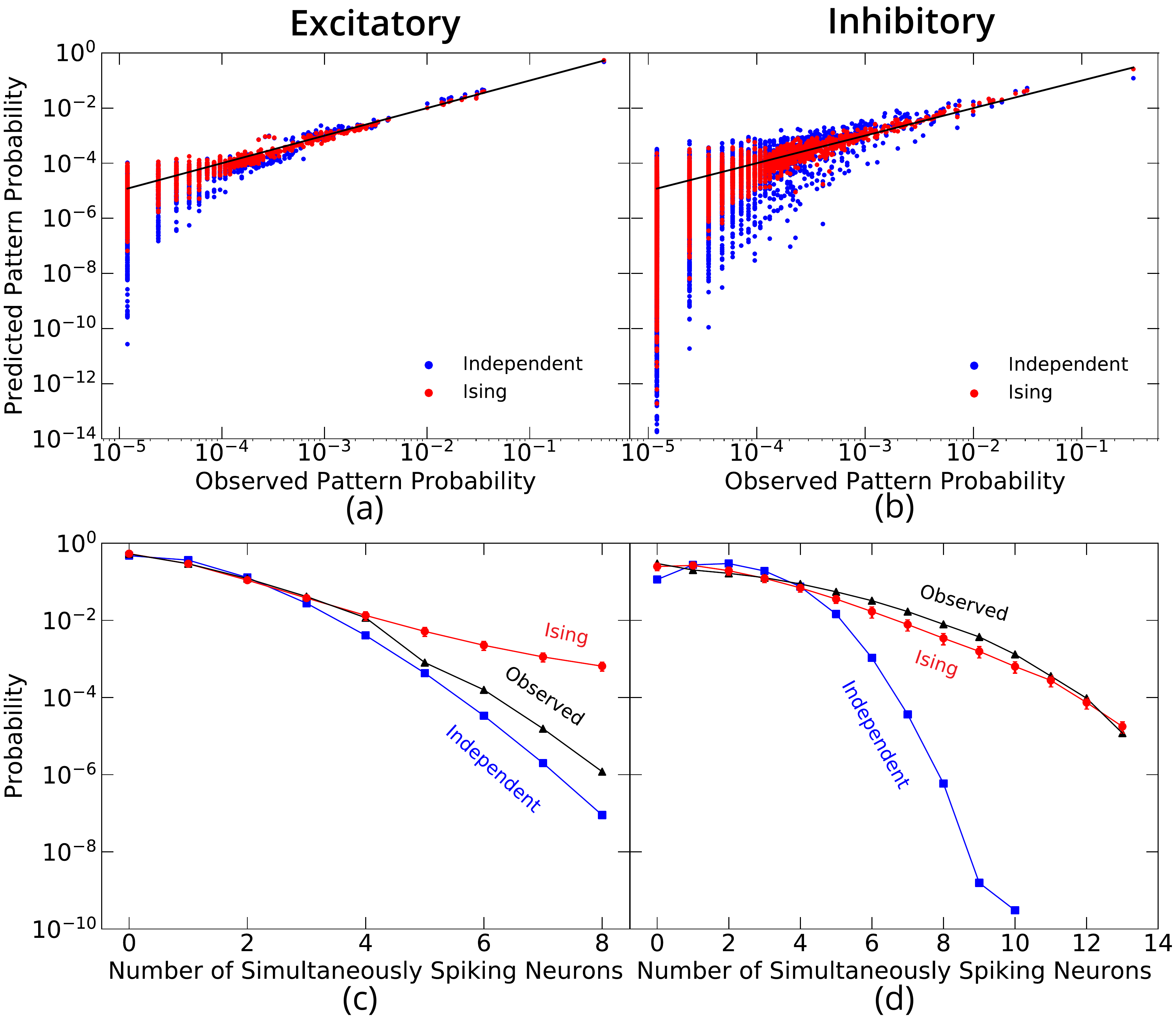}
\caption{\label{fig:EvsI} Difference between E- and I-neurons. The probability of occurrence of each firing pattern predicted by the maximum entropy model ($P^{(1)}(\sig, \h)$ or $P^{(2)}(\sig, \thet)$) is plotted against the observed pattern frequency from measured data ($P^{(N)}(\sig)$) for a network of (a) $14$ E-neurons, (b) $23$ I-neurons. The black line corresponds to prediction matching observation. The predicted and observed distribution of the number of simultaneously spiking neurons in each time bin for a network of (c) $14$ E-neurons, (d) $23$ I-neurons. Error bars are not symmetrical because of the logarithmic scale. All plots are for the deep sleep state, but similar results hold across all states and time binnings.}
\end{figure*}

\subsection{Collective behavior}
\label{sec:collective}

The first success of the pairwise model applied to reliable neurons can be seen when looking at the collective behavior in the network. Panel (a) of Figure~\ref{fig:good_model} shows the probabilities that the independent and Ising models assign to each spiking pattern. The Ising model is seen to significantly outperform the independent model, accurately predicting the observed pattern frequencies, except for the most rarely encountered patterns. Panel (b) of Figure~\ref{fig:good_model} shows the spike synchrony, defined as the probability that a given number of neurons spike within the same time window, revealing that the independent model strongly underpredicts events with many synchronous spikes. The Ising model is seen to perform dramatically better.

\subsection{Information-theoretic quantities}
\label{sec:info}

In addition to Figure~\ref{fig:good_model}, we can quantify the success of the Ising model by computing the information-theoretic quantities introduced in Section~\ref{sec:thermo} for different sleep states and subsets of neuron types. The results are summarized in Table~\ref{table:info}.

Recall that each firing pattern can be viewed as a vector of $N$ bits that specifies which of the $N$ neurons fired during a given time interval. Since the entropy $S_i$ can be interpreted as the number of bits required to describe a typical pattern drawn from the probability distribution $P^{(i)}$, we expect $S_i = N$ if all neurons randomly fired or remained silent with equal probability. However, since neurons are mostly inactive, the entropy is seen to be significantly smaller ($S_1/N\sim 0.3-0.4$) for the independent model that accounts for these firing rates, and even smaller ($S_2<S_1$) for the Ising model that also incorporates predictable regularities due to pairwise interactions. The entropy is the smallest ($S_N<S_2$) for the observed distribution, which includes regularities not captures by the Ising model.

The multi-information ratio $I_2/I_N = (S_1 - S_2)/(S_1 - S_N)$ (see Eq.~\ref{MultiInfoEq}) is listed in the last column of Table~\ref{table:info}. It reveals that the Ising model captures the majority of the correlative structure in the data: in transitioning from the independent model to the observed probability distribution for neuron firing patterns, the Ising model accounts for $80\% - 95\%$ of the correlations, depending on sleep state and neuron type.

Lastly, for a given model $P^{(i)}$, the Kullback-Leibler (KL) divergence $\KLD{P^{(i)}}{P^{(N)}}$ can be interpreted as the number of extra bits required to describe the observed neuron firing patterns simply because the model does not fit the data perfectly. Thus taking the KL divergence as a measure of how poorly a model fits the observed probability distribution of spiking patterns, Table~\ref{table:info} shows that the Ising model outperforms the independent model for all the nine cases listed, typically by a factor greater than three. We also see that excitatory neurons are the easiest to predict, and that the more awake the patient is, the more accurate the models become. 

\subsection{Excitatory and inhibitory neurons}

We now turn to a central question of this paper: identifying differences between inhibitory and excitatory neurons. We begin by investigating the interactions between these two neuron types. Figure~\ref{fig:EI} shows the couplings $J_{ij}$ between neurons of both types, revealing an interesting pattern:
the couplings among I-neurons and among E-neurons are almost all positive, whereas the couplings between I- and E-neurons are often negative, reflecting the ability of I-neurons to inhibit E-neurons. 

To further uncover the differences between inhibitory and excitatory neurons, we fit Ising models separately for each neuron type.
Figure~\ref{fig:EvsI} shows that the Ising model works fairly well for the I-neurons alone, but fails completely for the E-neurons alone, dramatically overpredicting how often neurons spike together. 

The success is arguably as noteworthy as the failure:
it is striking that the Ising model works as well as it does even though the couplings of our observed neurons to the other roughly $10^{11}$ neurons in the brain are completely ignored. One interpretation of this success is the well-known hypothesis that collective neural dynamics occurs on a low-dimensional submanifold \cite{stefanescu2008low}, such that observing a modest number of neurons suffices for determining a state's location in this submanifold.

Within the context of this interpretation, Figure~\ref{fig:good_model} illustrates how a subsystem of merely $19$ reliable neurons suffice to capture enough information about their surrounding dynamics to be accurately modeled as an isolated system, as long as both excitatory and inhibitory neurons are included. On the other hand, if the  inhibitory  effects  of  I-neurons  are  not  modeled, then  the synchrony among E-neurons is dramatically overestimated (see Panel (c) of Figure~\ref{fig:EvsI}). Conversely, there is a slight hint that if the excitatory effects of E-neurons are not modeled, then the synchrony among I-neurons is slightly, but consistently, underestimated (see Panel (d) of Figure~\ref{fig:EvsI}). In other words, accurate modeling of the joint system of E- and I-neurons requires sampling from both of its two distinct sub-populations.

\subsection{Sleep states}
\label{sec:sleep}

\begin{figure}
\centering
\includegraphics[width=\columnwidth]{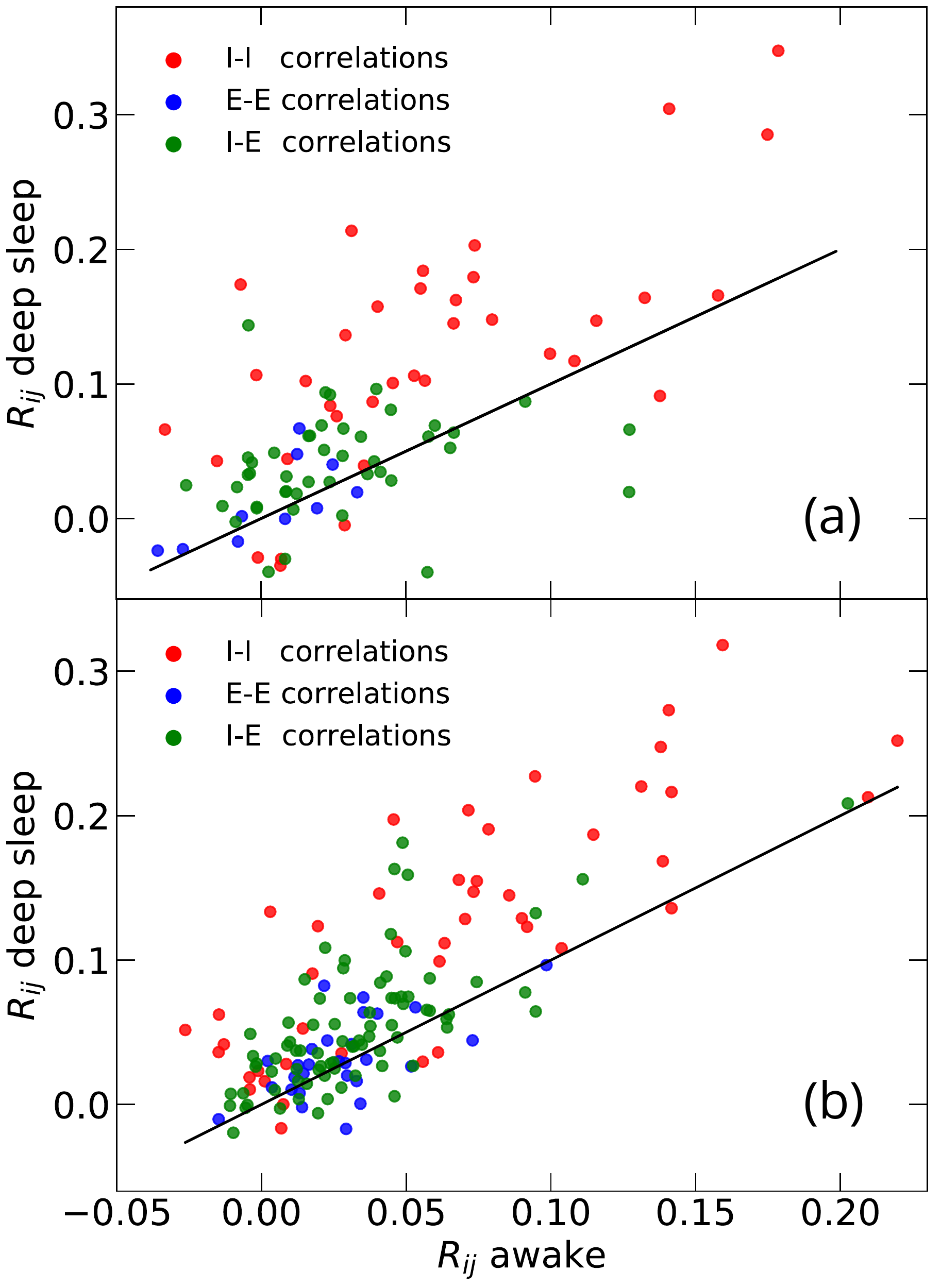}
\caption{\label{fig:RvsR} Pearson correlation coefficients $R_{ij}$ between the same set of (a) $14$ (patient 1) (b) $18$ (patient 2) reliable neurons during awake and deep sleep states. The coefficients $R_{ij}$ are labeled based on the type of neurons $i$ and $j$. The correlations are different during the two states, with I-I correlations being predominantly larger during deep sleep.}
\end{figure}

Now that we have described the behavior of inhibitory and excitatory neurons that is common to all sleep states, we proceed to investigate the distinctions that appear between E and I neurons in different states. It is well-known that the spiking frequencies of many neurons vary strongly between sleep states and our data confirms this fact \cite{Amzica1998slow,Destexhe1999sleepwake,Steriade2001wakesleep,Vyazovskiy2009firing,Peyrache2012spatiotemporal}. Additionally, we show that such a dependence on sleep states applies not only to the first moments $\m$ (corresponding to firing frequencies), but also to second moments $\C$. Figure~\ref{fig:RvsR} compares the Pearson correlation coefficients $R_{ij}\equiv C_{ij}/[C_{ii}C_{jj}]^{1/2}$ between all neuron pairs. Note that we only study the neurons that are reliable in both states and split the correlation coefficients according to neuron types. Noticeably, most dots lie far from the diagonal line, meaning that the corresponding correlations are different during wakefulness and sleep. For both patients, we observe that I-I correlations are slightly higher during deep sleep, which is indicated by their points lying predominantly above the diagonal line. This means that the network of inhibitory neurons is more internally correlated during sleep, perhaps reflecting the familiar slow-wave coherent oscillations associated with deep sleep.

\begin{figure}
\centering
\includegraphics[width=\columnwidth]{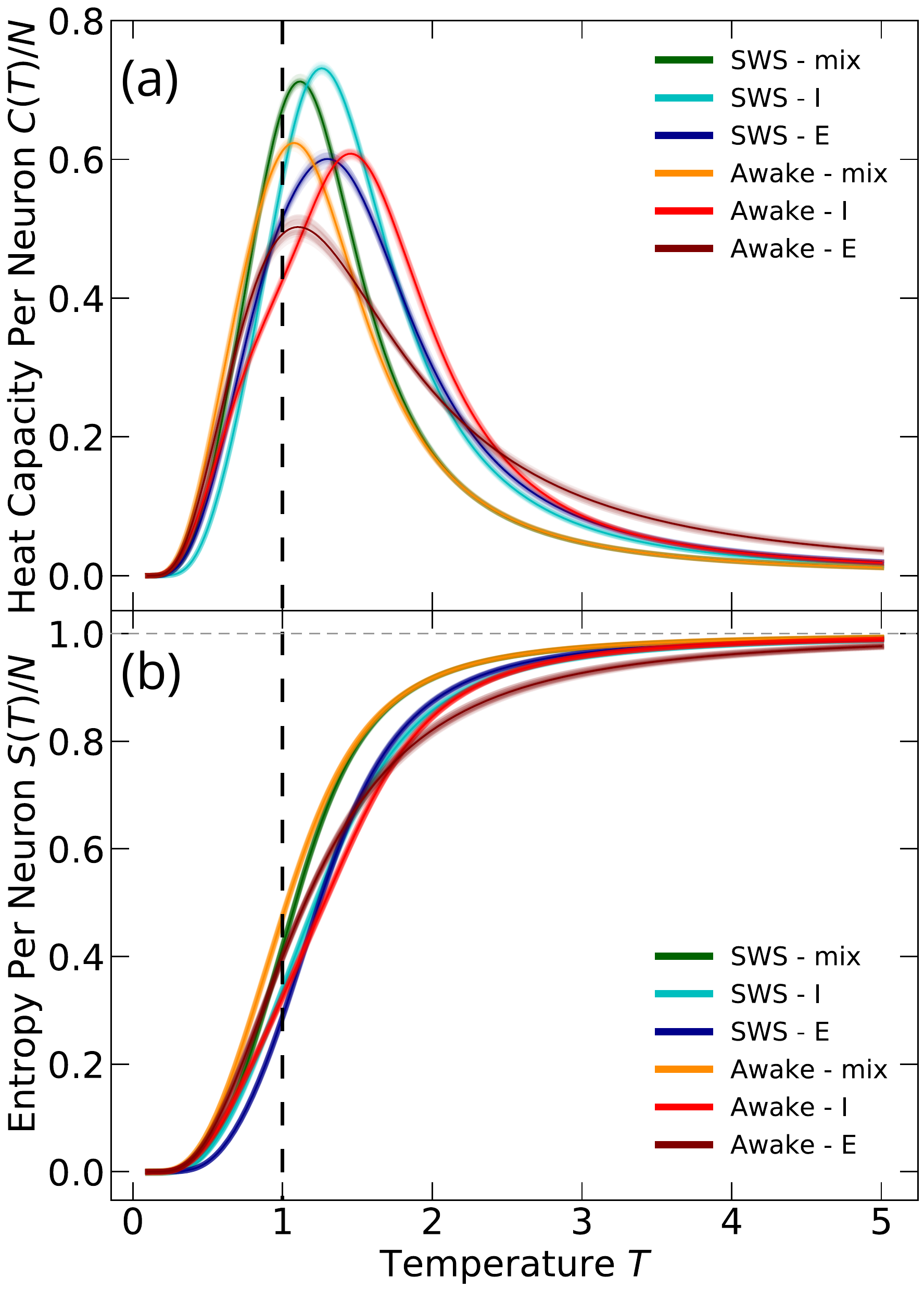}
\caption{\label{fig:heat} Heat capacity and entropy in maximum entropy models. (a) Heat capacity per neuron $C(T)/N$ as a function of temperature for different types of neurons in the awake and SWS states. The heat capacity peaks around $T = 1$ (vertical dotted line). (b) Entropy per neuron $S(T)/N$ as a function of temperature for different types of neurons in the awake and SWS states. The entropy experiences a jump around $T = 1$ and saturates to its maximum value $S(T)/N = 1$ (horizontal dotted line) at high temperatures. 
Both panels indicate signatures of criticality. Errors are displayed as shaded regions around the main lines. Similar results hold across all time bins.}
\end{figure}

\subsection{Thermodynamic quantities}
\label{sec:thermo_q}

Finally, we study the thermodynamic properties of our models. Figure~\ref{fig:heat} shows the heat capacity and entropy as a function of temperature for inhibitory, excitatory, and mixed networks of neurons in different sleep states. It is noteworthy that these thermodynamic curves look qualitatively similar across neuron types and sleep states despite the strong dependence on sleep state that we saw above at the level of individual neurons. We see that for all networks, the heat capacity peaks around the operating temperature $T = 1$, while the entropy exhibits a jump near that same temperature. This behavior is reminiscent of a phase transition, in which the system goes from an ordered, low-entropy state, to a disordered, high-entropy state \cite{kardar2007statistical}. At high temperatures, where all spiking patterns have the same Boltzmann weight, the entropy approaches its maximum $S(T) = N$. The peak in the heat capacity curves becomes sharper for larger system size $N$, which is consistent with finite-size scaling. In all the networks studied, the heat capacity peak is found to be at a slightly higher temperature than the operating point $T = 1$. Moreover, these signatures of criticality are robust to changes in parameters $\thet$, as indicated by the small size of our error bars. The question of interpreting these criticality hints is subtle and we will return to it in Section~\ref{sec:criticality}.
\section{Discussion}
\label{sec:discussion}

In this paper, we have introduced a statistically rigorous and computationally efficient method for inferring an Ising model of the spiking activity of neural networks, and have applied it to excitatory and inhibitory human cortical neurons during the wake-sleep cycle. Our method provides accurate uncertainty estimates for all model parameters and derived quantities, as described in Appendix~\ref{sec:error}, and remains tractable for large ($N\approx100$) neuronal networks. This further improves previous work where parameters uncertainty was estimated either  by repeatedly running the inference algorithm for different neuron subsets \cite{schneidman2006weak, tkacik2009spin,tkavcik2014searching, tkavcik2015thermodynamics, ganmor2011sparse,ganmor2011architecture,mora2015dynamical} or not at all \cite{shlens2006structure,shlens2009structure,nghiem2018maximum}; the former approach may suffer from the population of neurons being inhomogeneous, consisting of different neuron types, being exposed to variable stimuli, or having drastically varying firing rates. 

\subsection{Modeling spiking behavior}

We found that the Ising model describes neuronal collective behavior much better than the independent model throughout the sleep cycle as long as both excitatory (E) and inhibitory (I) neurons are modeled (Figure~\ref{fig:good_model}). These observations are consistent across sleep states and patients. We found that the pairwise correlations in the Ising model accounted for $80\%-95\%$ of all correlations in the data and that the Ising model yielded a KL divergence more than three times smaller than that of the independent model (Table~\ref{table:info}). 

By modeling inhibitory and excitatory networks separately, we found that accurately predicting neuron synchrony requires sampling both E- and I-neurons. If the inhibitory effects of I-neurons were ignored, then synchrony among E-neurons was dramatically overestimated (Panel (c) of Figure~\ref{fig:EvsI}). In contrast, the I-neurons could be fairly accurately modeled on their own, although ignoring the excitatory effects of the E-neurons caused a slight underestimation of their synchronous activity (Panel (d) of Figure~\ref{fig:EvsI}). The inhibitory effect of I-neurons on E-neurons was also reflected by negative pairwise couplings $J_{ij}$ between the two neuron types  (Figure~\ref{fig:EI}).

A recent study \cite{nghiem2018maximum} also examined excitatory and inhibitory neurons in the human cortex during wakefulness and deep sleep, reporting that the spiking activity is dominated by pairwise interactions during wakefulness but is population-wide during deep sleep \cite{nghiem2018maximum}, to an extent not captured by the Ising model for inhibitory neurons. While we found I-neurons to be accurately fit by the Ising model, we reproduced the conclusion of \cite{nghiem2018maximum} that inhibitory neurons have higher average correlation during sleep, by analyzing the intra-class (I-I, E-E) and inter-class (E-I) interactions  (Figure~\ref{fig:RvsR}). Moreover, we found this higher I-I correlation to be more pronounced in deep sleep than in light sleep. 

These findings further complement prior observed differences between excitatory and inhibitory neurons in the wake-sleep cycle. It has been shown that the cortical neurons manifest an overall multiscale balance, i.e. ensemble excitation and inhibition co-fluctuate, a property that is observed across multiple timescale and involves transient deviations from the absolute balance that are more prominent during deep sleep \cite{Dehghani2016balance}. In addition, excitatory neurons show a tendency to be active more focally, manifesting a distance-dependent decay in their correlation, while in contrast, inhibitory neurons show a more robust correlation within the span of a cortical column \cite{Peyrache2012spatiotemporal}. These collective features are well matched with the envisioned significant role of the inhibitory neurons in gating information \cite{Steriade1993slow,steriade2003neuronal} and in regulating oscillations including slow-wave sleep \cite{Peyrache2012spatiotemporal, LVQ2016oscillation}. 

\subsection{Is the cortex critical?}
\label{sec:criticality}

Whether the collective neural activity is optimized to operate at a critical point \cite{tkavcik2015thermodynamics,Beggs2003avalanche,Chialvo2010complex}
or reflects an ``asynchronous irregular'' (AI) regime \cite{Ecker2010decorrelated,Softky1993irregular} (as a result of irregular firing with weak mean correlations despite substantial shared input) remains a controversial topic. A number of studies have used thermodynamic-based measures of population activity, such as the divergence of heat capacity and the power-law scaling of neural activity \cite{mora2011biological,yu2013universal,tkavcik2015thermodynamics,mora2015dynamical,ioffe2017structured}, as evidence in support of criticality. Each of these hypotheses has important implications for neural coding. On one hand, the critical regime provides optimal information retention \cite{Haldeman2005branching}, information integration \cite{tegmark2015consciousness},  and maximum sensitivity to input variations \cite{Kinouchi2006optimal}. On the other hand, the AI regime reflects a dynamic balance state of excitatory and inhibitory fluctuations \cite{Renart2010ai,vanVreeswijk1996chaos}, providing non-redundant fast network responses \cite{vanVreeswijk1996chaos}. 

We tested for criticality by measuring the heat capacity and entropy as a function of temperature. We found that although the activity and correlation between individual neurons varied dramatically with sleep state (Figure~\ref{fig:RvsR}), the thermodynamic quantities remained qualitatively unchanged, with the heat capacity $C(T)$ peaking just after $T=1$ (Panel (a) of Figure~\ref{fig:heat}) and entropy $S(T)$ experiencing a significant jump at $T\approx1$ (Panel (b) of Figure~\ref{fig:heat}). These characteristics may suggest long-range correlation at the critical temperature, allowing the neurons to coordinate across extended distances. As a consequence, the divergence of response functions would maximize the sensitivity of the system to the stimuli. In addition, the jump in entropy near the critical point implies a significant increase in the number of effective spiking patterns available to the system, thus allowing it to exhibit a larger spectrum of possible responses. 

However, we wish to draw attention to experimental and theoretical evidence for why we should take these criticality indications with a grain of salt.  
Experimentally, previous analysis of our data set have shown that across the wake-sleep cycle, human (as well as monkey and cat) cortical neurons do not show power-law scaling and are better fit with a multi-exponential model, suggesting that the underlying dynamics mirror the interaction of excitation and inhibition at multiple timescales \cite{Dehghani2012avalanche,Dehghani2016balance}. 

Turning to theoretical evidence, shared input to the network \cite{schwab2014zipf,aitchison2016zipf}, higher-order couplings \cite{Macke2011input}, and randomly sub-sampled data \cite{nonnenmacher2017signatures,Priesemann2014subcritical} can all masquerade as signatures of criticality. In fact, networks in self-sustained irregular regimes away from criticality can still manifest universal scaling functions \cite{Touboul2017criticality}. Our data, and essentially any {\it in vivo} multielectrode measurements, dramatically subsample the neural activity and only provide partial measurements to estimate correlations, and leave us with latent variables that can only be inferred indirectly. 

Finally, it has been suggested that the alleged criticality of the maximum entropy models can be a consequence of the inference procedure \cite{mastromatteo2011criticality,marsili2013sampling,tyrcha2013effect,haimovici2015criticality}. The basic argument is that the Ising model only supports long-range correlations when it is near-critical, such that if the data contains long-range correlations, then the Ising model that best fits these correlations is likely to be near-critical. The fact that we observe signatures of criticality independent of sleep state (awake, light sleep, and deep sleep), neuron type (E, I, and mixed), and timescale ($20$, $50$, and $100$ ms) thus suggests long-range correlations in all states that may or may not be due to critical behavior. A more detailed analysis of spatial and temporal correlations is required in order to settle the criticality controversy.

\subsection{Outlook}

To shed further light on the workings of biological neural networks, there are many opportunities to  improve the method we have presented. 
The Ising model is simply the maximum-entropy model that matches all firing rates and equal-time two-point correlations.
Therefore, straightforward generalizations involve including correlations between different times and including three-point functions and higher-order correlations.
 
The addition of higher-order moments as constraints has the potential to improve the model \cite{amari2003synchronous, montani2009impact, montani2013statistical,koster2014modeling,shimazaki2015simultaneous,leen2015simple}, especially since input nonlinearities have been argued to affect beyond-pairwise correlations \cite{Josic2009stimulus,zylberberg2015input}. However, even the inclusion of three-point functions already results in a model with $O(N^3)$ parameters, which requires significantly more data points to avoid overfitting. Moreover, adding higher-order correlations renders the optimization problem even harder since the energy landscape will likely have even more local minima. Several promising methods have been introduced for capturing beyond-pairwise correlations, \cite{Macke2011input,Gardella2016method,o2017population}, but important work remains to be done to avoid overfitting and make the computation of the model parameters and their uncertainties computationally tractable. 

Spatiotemporal extensions of the maximum entropy model to include correlations between different times have also been pursued in the past \cite{tang2008maximum,marre2009prediction, vasquez2012gibbs}. However, there are still plenty of interesting questions to study using these models and many valuable opportunities for improving their reliability, uncertainty estimation, and computational efficiency. For example, a spatiotemporal extension of our analysis should be able to explicitly quantify which neurons are exciting or inhibiting others at later times. 

There are also ample opportunities to tackle the above-mentioned subsampling limitation, i.e.,  that current neuron data tends be recorded from merely a minuscule fraction of all neurons. If it is correct that the relevant dynamics occur on a low-dimensional submanifold, then there is great value in further experimental and theoretical work to determine its dimensionality (and how many neurons suffice for capturing said dynamics). 

Fortunately, rapid technological progress is enabling simultaneous high-quality recordings from ever-larger numbers of neurons. This will produce a gold mine of data that can be tested using our method and further improvements thereof to deepen our understanding of biological neural network dynamics.

\begin{acknowledgments}
The authors would like to thank James Crutchfield, Cina Aghamohammadi, and Joshua Deutsch for helpful discussions, and the Center for Brains, Minds, and Machines (CBMM) for hospitality. This work was supported by NSF grant 1734870, the Foundational Questions Institute, and the Rothberg Family Fund for Cognitive Science. CZ was supported by the Whiteman Fellowship.
\end{acknowledgments}

\appendix
\section{Patients and Recordings}
\label{sec:data-appendix}
For this study, we used 4 patient/session multielectrode temporal cortex recordings, each lasting 12 hours, including overnight sleep. The recordings were done using silicon-based NeuroProbe (from BlackRock Microsystems Inc.), composed of a $10\cross10$ 2D array of micro-electrodes, each $1$ mm thick, separated by a $400$ $\mu$m spacing \cite{Campbell_1991mea,Jones1992mea}. Four corner electrodes were used for grounding the electronics, resulting in $96$ functional electrodes sampling the data at $30$ kHz. All patients had focal epilepsy as confirmed by postoperative histology. Based on the post-excision histological exams, electrode tips reached the layer II/III of the middle temporal gyrus. The array implantation and surgical excision were performed for medical purposes according to IRB approved experiments. Patient multielectrode recording experiment approval was granted by the Institutional Review Boards of Massachusetts General Hospital, and Brigham $\&$ Women's Hospital. Sleep staging was performed based on video monitoring and a combination of scalp EEG and intracranial EEG recordings simultaneously acquired with the multielectorde array system. Sleep-wake categories were assigned as awake, light sleep, deep sleep (SWS), and rapid eye movement (REM). Due to the short duration of REM sleep, we did not use the REM sleep data in our study. For more details on patients and recordings, see the methods and supplementary material of prior publications \cite{Peyrache2012spatiotemporal,LVQ2016oscillation}.

\section{Spike Sorting and Cell Categorization}
\label{sec:spike-appendix}
After thresholding the raw data for spike detection, the selected spikes were sorted offline using the first three principal components of spike waveforms from each electrode. This step was then followed by an automated clustering using expectation-maximization (EM). The overall morpho-functional characteristics of the spike waveform and putative mono-synaptic connections provided the final two cell classes of excitatory (E) and inhibitory (I). First, each cell's average spike waveforms was used to measure a variety of features, such as half-width of the positive peak, half-width of the negative peak, interval between negative and positive peaks (valley-to-peak), and the ratio of the negative to positive peak amplitude. Based on a K-means algorithm, these features were used to categorize the cells based on the morphology of the spike waveform of Fast-Spiking (FS, putative inhibitory) and Regular-Spiking (RS, putative excitatory) \cite{McCormick1985cells,Bartho2004fsrs}. The FS and RS functional labels were later tested and verified by the cross-correlograms  interactions indicative of putative monosynaptic connections \citep{Fujisawa2008mono}. Only the verified categorized cells were used in the analyses. This resulted in $4$ data sets, each with $92$, $80$, $36$, and $30$ neurons respectively. More details about spike sorting and cell classification are provided in the methods and supplementary material of \citep{Peyrache2012spatiotemporal,Dehghani2016balance}.

\section{Learning the parameters of the maximum entropy model}
\label{sec:learning}

To solve the inverse Ising problem, we must find the parameters $\thet$ such that the Boltzmann distribution $P^{(2)}(\sig, \thet)$ is as close as possible to the observed distribution $P^{(N)}(\sig)$ over the data set $\X = \{\sig^{(1)}, \sig^{(2)}, \ldots, \sig^{(M)}\}$. 
More formally, this implies maximizing the likelihood
\begin{equation}
L(\X, \thet) = \prod_{k=1}^M \dfrac{e^{- H(\sig^{(k)},  \thet)}}{Z(\thet)} = \dfrac{e^{- \sum_{k=1}^MH(\sig^{(k)}, \thet)}}{Z^M(\thet)}.
\end{equation}
that the data is produced by our model.
In practice, it is often more convenient to consider the log-likelihood 

\begin{align}
\label{eq:ll}
\log L(\X, \thet) &= - \sum_{k=1}^MH(\sig^{(k)}, \thet) - M\log Z(\thet) \nonumber\\
&=  \sum_{k=1}^M\left(\h^T \sig^{(k)} + {\sig^{(k)}}^T \J \sig^{(k)}\right) - M\log Z(\thet) \nonumber\\
&= M \sum_{i = 1}^Nh_im_i + M\sum_{i, j = 1}^NJ_{ij}Q_{ij} - M\log Z(\thet) \nonumber\\
&= M \left[\h^T\m + \Tr\left(\J\Q\right) - \log Z(\thet)\right].
\end{align}
Notice that the log-likelihood depends only on one- and two-point correlation functions, and not on the entire data set. This is consistent with our expectation that $\m$ and $\Q$ provide sufficient information to learn the pairwise model.  

The log-likelihood in Eq.~\eqref{eq:ll} can be maximized by using optimization algorithms, such as gradient descent. The gradient of the log-likelihood is straightforward to compute \cite{nguyen2017inverse} and is given by

\begin{equation}
\dfrac{\partial \log L(\X, \thet)}{\partial h_i} = M(m_i - m_i(\thet)),
\end{equation}

\begin{equation}
\dfrac{\partial \log L(\X, \thet)}{\partial J_{ij}} = M(Q_{ij} - Q_{ij}(\thet)),
\end{equation}
where $\m(\thet)$ and $\Q(\thet)$ denote the first and second moments predicted by the Ising model using
Eqs~\eqref{eq:m} and~\eqref{eq:C}.
The gradient descent update rule at each iteration is simply
\begin{equation}
\h \leftarrow \h + \eta[\m-\m(\thet)],
\end{equation}

\begin{equation}
\J \leftarrow \J + \eta[\Q-\Q(\thet)],
\end{equation}
where we have absorbed $M$ into the learning rate $\eta$. 
Once we reach the maximum log-likelihood, the gradient and hence these updates will vanish, implying that our model will match the first and second moments of the observed distribution, as expected for a pairwise maximum entropy model. 

In order to compute the average values $\m(\thet)$ and $\Q(\thet)$ on the right-hand side of the equations above, we need to sum over all $2^N$ possible configurations of the system. For large system sizes ($N > 30$), this computation becomes intractable. We therefore use Markov Chain Monte Carlo (MCMC) \cite{metropolis1953equation,hastings1970monte} with the usual Metropolis-Hastings sampling procedure to generate representative samples from $P^{(2)}(\sig, \thet)$ and estimate $\m(\thet)$ and $\Q(\thet)$ using Eqs.~\eqref{eq:m} and ~\eqref{eq:C}. In practice, we draw $10^6$ samples and set $\eta = 0.01$. 

At each iteration, we need to assess how closely our model matches the data. Ideally, we would compute the log-likelihood, but estimating the partition function takes an exponential amount of time. Therefore, we use the root-mean-squared-error (RMSE) between the predicted and measured one- and two-point functions as a proxy for monitoring the convergence of the optimization algorithm
\begin{equation}
\ell = \sqrt{\dfrac{1}{N}\sum_{i=1}^N(m_i-m_i(\thet))^2 + \dfrac{1}{N^2}\sum_{i, j = 1}^N(Q_{ij}-Q_{ij}(\thet))^2},
\end{equation}
which is simply the likelihood gradient magnitude $|\nabla_\thet L|$ except for two normalization factors.
Our success criterion for terminating the optimization procedure is $\ell < 0.001$. 

\section{Estimating uncertainties on the parameters of the Ising model}
\label{sec:error}

\subsection{MCMC on parameter space}
\label{sec:mcmc}

Suppose that our optimization routine, described in Appendix~\ref{sec:learning}, converged to $\thet_* = (\h_*, \J_*)$, which is the maximum likelihood estimate of our model parameters given a data set $\X$. We now start at $\thet_*$ and perform a random walk in the space of parameters $\thet$ using MCMC with a Gaussian proposal distribution and assuming an uniform prior. The algorithm is as follows:

\begin{enumerate}
   \item Initiate the MCMC chain with $\thet_0 = \thet_*$. 
   \item At each iteration $t$:
   \begin{itemize}
     \item Draw a candidate step $\mathbf{s}_t \in \R^{N(N+1)/2}$ according to the multivariate Gaussian distribution $\mathcal{N}(0, \Sig)$. The proposed parameter vector is $\thet' = \thet_{t-1} + \mathbf{s}_t$. 
     \item Accept the proposed move with probability 
     
	\begin{align}
     &p_{accept} = \min\left[1, \dfrac{L(\X, \thet')}{L(\X, \thet_{t-1})} \right] \nonumber\\
     &= \min\left[1, e^{M\left(\left(\h'-\h_{t-1}\right)^T\m + \Tr\left((\J'-\J_{t-1})\Q\right) + \log(\frac{Z(\thet_{t-1})}{Z(\thet')}) \right)}\right].
	\end{align}
     
   \end{itemize}
   \item After generating $k$ representative samples $\{\thet_1, \ldots, \thet_k\}$, we can use them to compute the mean and standard deviation on $\thet$, thus quantifying the uncertainty for the Ising model parameters.

\end{enumerate}

\subsection{Adaptive MCMC}
\label{sec:adaptive}

One important technicality is choosing the step covariance matrix $\Sig$. If $\Sig$ is too small, then most of our proposals will be accepted but we will not get to explore much of the parameter space. If $\Sig$ is too large, our steps will be very big and most of our proposals will be rejected. Moreover, we do not know how large $\Sig$ has to be in each dimension. 

A standard way to deal with these problems is to employ adaptive MCMC techniques, which use the history of previous samples to update the proposal distribution dynamically at each step \cite{andrieu2008tutorial}. We can update $\Sig$ using either a fixed number of previous states, as is the case for Adaptive Proposal (AP) MCMC \cite{haario1999adaptive}, or the whole chain generated so far, as in Adaptive Metropolis (AM) MCMC \cite{haario2001adaptive}. Updating the proposal distribution during our algorithm gives us a better chance at convergence. However, the adaptive algorithms lead to a stochastic process that is clearly no longer Markovian. Therefore, adaptive schemes may converge to incorrect distributions, since the standard ergodicity result no longer applies to non-Markovian processes. This is indeed the case for the AP algorithm, whereas AM has proven to be ergodic \cite{haario2001adaptive}. Therefore, we  use the AM algorithm in this work. 

The covariance matrix $\Sig_t$ at each iteration $t$ is chosen according to 

\begin{equation}
    \Sig_t= 
\begin{cases}
    \Sig_0 & \text{if } t\leq t_0\\
    \lambda_d \mathrm{Cov}(\thet_1, \ldots, \thet_{t-1}) & \text{if } t > t_0
\end{cases}
\end{equation}
where $t_0$ is the initial period after which adaptation begins. Typically, we set $t_0 = 500$ and run the simulation for $10,000$ steps. The scaling parameter $\lambda_d = 2.4^2 / d$ depends only on the dimension of the parameter space $d = N(N-1)/2$ and is chosen so that it optimizes the mixing properties of the random walk in the case of Gaussian proposal and target distributions \cite{gelman1996efficient}. As an initial guess for the covariance matrix we choose the identity $\Sig_0 = \alpha \mathbf{I}_d$, scaled down by a factor $\alpha = 10^{-5}$ such that approximately half of the proposed steps get accepted during MCMC.

\subsection{Approximating ratios of partition functions}
\label{sec:approx}

The algorithm presented above relies on computing the ratio of partition functions in its second step. This can be performed exactly for small systems, where we can directly compute the partition function. However, it becomes unfeasible for larger systems, including those with $N\approx100$ which we are interested in. Therefore we need a way to approximate the ratio of partition functions. 
 
Although there are a few methods for directly estimating the partition function, such as the Wang-Landau algorithm (see Appendix~\ref{sec:wang-landau}), they can be computationally expensive to run at each MCMC step and for now we are more interested in approximating the ratio of partition functions $Z(\thet')/Z(\thet)$, rather than the partition functions themselves. We can re-write the ratio as follows:

\begin{align}
\label{eq:ratio}
\dfrac{Z(\thet')}{Z(\thet)} &= \dfrac{1}{Z(\thet)}\sum_{\sig}e^{-H(\sig, \thet')} = \dfrac{1}{Z(\thet)}\sum_{\sig} e^{-H(\sig, \thet')}  \dfrac{P^{(2)}(\sig, \thet)}{P^{(2)}(\sig, \thet)} \nonumber \\
&= \sum_{\sig} \dfrac{e^{-H(\sig, \thet')}}{e^{-H(\sig, \thet)}}P^{(2)}(\sig, \thet) = \left\langle e^{-H(\sig, \thet' - \thet)} \right \rangle_\thet.
\end{align}

Now we can approximate the right-hand side by drawing Monte Carlo samples from the model with parameters $\thet$. If we draw $M$ such samples $\sig^{(k)}$, then

\begin{equation}
\dfrac{Z(\thet')}{Z(\thet)} \approx \dfrac{1}{M}\sum_{k=1}^M e^{-H(\sig^{(k)}, \thet' - \thet)},
\label{eq:Z_approx}
\end{equation}
where the approximation sign was used to denote the fact that the two equations are equal only in the limit of infinite samples $M$. In practice, the right-hand side of Eq.\eqref{eq:Z_approx} is a good approximation only when the distribution $P^{(2)}(\sig, \thet)$ is close to $P^{(2)}(\sig, \thet')$. If the two probability distributions are not close to each other, then most samples drawn from the $\thet$-model will have a very low probability under the $\thet'$-model and therefore make a negligible contribution to the sum in Eq.~\eqref{eq:Z_approx}. 

Fortunately, since we will be computing ratios of partition functions for model parameters drawn on consecutive iterations, $\thet_{t-1}$ and $\thet_t$, and since our covariance matrix $\Sig$ tends to be very small, it is reasonable to assume that $\thet_{t-1}$ and $\thet_t$ are close enough to each other, such that Eq.~\eqref{eq:Z_approx} is applicable with relatively few ($M\approx10^5$) samples. Therefore, the runtime at each MCMC iteration $t$ will mostly consist of drawing $M$ Monte Carlo samples from our previous model $\thet_{t-1}$.


\section{The Wang-Landau algorithm}
\label{sec:wang-landau}

The main idea behind the Wang-Landau algorithm \cite{wang2001efficient} is to 
directly estimate the density of states $g(E)$, i.e. the number of states (spin configurations) with a given energy $E$. Once we know the density of states, we can compute the partition function by grouping the terms in the sum by energy

\begin{equation}
Z(\thet) = \sum_{\sig'}e^{-\beta H(\sig', \thet)} = \sum_E g(E) e^{-E/T}.
\end{equation}
Other important thermodynamic quantities, such as the average energy, heat capacity, and entropy, follow from the partition function:

\begin{equation}
\langle E\rangle_\thet = \dfrac{1}{Z(\thet)}\sum_E g(E) E e^{-E/T},
\end{equation}

\begin{equation}
C(T) = \dfrac{\langle E^2\rangle_\thet - \langle E\rangle_\thet^2}{T^2} ,
\end{equation}

\begin{equation}
S(T) = \dfrac{\langle E\rangle_\thet - F}{T} = \dfrac{\langle E\rangle_\thet}{T} + \ln Z(\thet),
\end{equation}
where $F = -T\ln Z(\thet)$ is the free energy. It is worth mentioning that $g(E)$ is independent of temperature. Hence we can compute the above quantities at any temperature $T$ without rerunning the algorithm.

In order to estimate the density of states, the Wang-Landau algorithm performs a random
walk in energy space and accepts the energy $E$ associated with each spin configuration with a certain probability, designed to encourage the exploration of states with different energies \cite{landau2004new}. For systems with continuous spectra, or systems with a lot of accessible energies (as is the case here), we begin by discretizing the spectrum into energy levels that are $\Delta$ apart. We assume that the energy spectrum is bounded, such that there is a finite number of energy levels. During the random walk, we keep a histogram $\rho(E)$ which is incremented each time we visit a state with energy $E$. The histogram has support only on the discretized, bounded spectrum. The random walk continues until the energy histogram $\rho(E)$ becomes flat, i.e. all its entries are within $20\%$ of the mean value $\langle \rho(E) \rangle$. 
We check the flatness of the histogram every $10,000$ steps.  

The procedure during the random walk is as follows \cite{landau2004new}:

\begin{enumerate}
   \item Initialize $g(E) = 1$ and $\rho(E) = 0$ for all energies $E$. 
   \item At each iteration $t$:
   \begin{itemize}
     \item Generate a new configuration by randomly flipping a spin $\sigma_i$. 
     \item Let $E_{t-1}$ and $E'$ be the energies of the previous and current configurations respectively. Accept the new configuration and set $E_t = E'$ with probability
     
     \begin{equation}
     p_{accept} = \min\left(1, \dfrac{g(E_{t-1})}{g(E')} \right).
     \end{equation}
     Otherwise, $E_t = E_{t-1}$.
     
     \item Update the density of states by a modification factor $f$, i.e. $g(E_t) \rightarrow fg(E_t)$.
     
   \end{itemize}
   \item If the histogram $\rho(E)$ is flat and $\ln f > \epsilon$, reduce the modification factor $f \rightarrow \sqrt{f}$, reset the histogram $\rho(E) = 0$, and proceed to step $2$.
\end{enumerate}
The modification factor $f$ controls how well we approximate $g(E)$. Therefore, periodically decreasing $f$ leads to a finer approximation of the density of states. Typical values for the parameters of the algorithm are $\Delta = 0.005$, $f = e = 2.7182$, and $\epsilon  = 10^{-11}$.

\bibliographystyle{apsrev4-1}
\bibliography{references}

\end{document}